**Thermodynamic Theory of Linear Optical and Electro-Optic Properties of Ferroelectrics**

*Aiden Ross\*, Mohamed S.M.M. Ali, Akash Saha, Rui Zu, Venkatraman Gopalan, Ismaila Dabo, Long-Qing Chen*

Department of Materials Science and Engineering and Materials Research Institute, The Pennsylvania State University, University Park, PA 16802

**Abstract:**

Ferroelectric materials underlie key optical technologies in optical communications, integrated optics and quantum computing. Yet, there is a lack of a consistent thermodynamic framework to predict the optical properties of ferroelectrics and the mutual connections among ferroelectric polarization, optical properties, and optical dispersion. For example, there is no existing thermodynamic model for establishing the relationship between the ferroelectric polarization and the optical properties in the visible spectrum. Here we present a thermodynamic theory of the linear optical and electro-optic properties of ferroelectrics by separating the lattice and electronic contributions to the total polarization. We introduce a biquadratic coupling between the lattice and electronic contributions validated by both first-principles calculations and experimental measurements. As an example, we derive the temperature and wavelength-dependent anisotropic optical properties of $BaTiO_3$, including the full linear optical dielectric tensor and the linear electro-optic (Pockels) effect through multiple ferroelectric phase transitions, which are in excellent agreement with existing experimental data and first principles calculations. This general framework incorporates essentially all optical properties of materials, including coupling between the ionic and electronic order parameters, as well as their dispersion and temperature dependence, and thus offers a powerful theoretical tool for analyzing light-matter interactions in ferroelectrics-based optical devices.

\* Email Address: amr8057@psu.edu

## I. INTRODUCTION

Understanding and controlling light-matter interactions underlies modern technologies ranging from computation to telecommunications. There has been significant recent interest in the electro-optic responses of ferroelectric materials, in which the coupling between the ferroelectric and optical properties leads to their substantial enhancements, making ferroelectrics a critical platform for photonic integrated circuits for high-speed photonic modulation [1], photonic neuromorphic computing [2], and quantum photonic computation. [3] For example, ferroelectric crystals such as $LiNbO_3$ currently power the internet through electro-optic modulators [4] and more recently, $BaTiO_3$ has attracted significant interest for quantum computing [3].

However, despite recent progress, there is still a lack of a general thermodynamic formulation to model and analyze the optical properties of ferroelectrics and their associated couplings to electric field, stress, and strain, which is critical for optimizing or enhancing their desired optical properties. For example, none of the current thermodynamic theories [5–7] incorporate the coupling between the ferroelectric order and the optical properties in the visible spectrum, and there is no theory which allows us to analyze the connections among the ferroelectric polarization, stress, and dispersion in the optical properties.

The thermodynamics of ferroelectrics have often been studied by employing the Landau–Ginzburg–Devonshire (LGD) theory to describe and relate the properties of various ferroelectric materials. It describes the change in free-energy density from the paraelectric to the ferroelectric state as a function of the ferroelectric polarization [8,9] and its couplings to strain and other order parameters. The second- and higher-order derivatives of the free energy density function with respect to the ferroelectric polarization and other order parameters or their thermodynamic conjugate variables describe the ferroelectric contributions to the tensorial ferroelectric and coupling properties. For example, the piezoelectric tensor, measuring the strength of the coupling between the ferroelectric polarization and stress, can be obtained from the second-order derivatives of the free energy density function with respect electric field and stress. Therefore, in principle, this free energy density function can be parameterized using the properties from experiments [10] and first principles calculations. [11] This free-energy density function serves as the basis for including the intrinsic thermodynamic properties in the phase-field method for modeling mesoscale ferroelectric domain structures and their associated properties, allowing for a predictive, multiscale materials modeling approach. [12]

The dielectric susceptibility determines the optical properties of materials. The ferroelectric contribution to the dielectric susceptibility can be calculated from the LGD free energy density function based on the susceptibility of the *total* ferroelectric polarization. However, for calculating the optical properties of ferroelectrics in the near-infrared to visible spectrum, corresponding to frequencies at which the ion cannot respond, using the total polarization is no longer suitable.

By considering the change in the dielectric susceptibility with an applied electric field, LGD theory has previously been employed to calculate the electro-optic effect in ferroelectrics in the visible spectrum. [5–7] However, this approach assumes that the susceptibility of the ferroelectric polarization is the main contribution to the total dielectric susceptibility. This assumption is only valid below the resonance frequency of the total ferroelectric polarization, which is typically in the microwave to terahertz frequency range, and hence it does not capture the relationship between the ferroelectric polarization and the optical properties in the visible spectrum. Therefore, there is no existing formulation of LGD theory that can accurately describe the optical properties of ferroelectrics in the near-infrared to visible spectrum.

To compute the optical properties in the near-infrared to visible wavelength, the connection between the electronic contribution to the dielectric susceptibility and the ferroelectric polarization must be included in a thermodynamic description. Previous models, primarily developed by Wemple and DiDomenico, utilized a polar-optic tensor to describe the optical properties of ferroelectrics [13–15]

$$\Delta \left(\frac{1}{n(\lambda)}\right)^2_{ij} = \Delta B_{ij}(\lambda) = g_{ijkl}(\lambda)P_k P_l. \tag{1}$$

where $n$ is the refractive index, $\lambda$ is the wavelength of light, $B_{ij}$ is the optical indicatrix, $g_{ijkl}$ is the polar optic tensor, $P_i$ is the ferroelectric polarization, and $\Delta$ represents the change with respect to the $P_i = 0$ state.

Using the polar-optic tensor, the total electro-optic effect can be calculated as the product of two parts: the change of the optical indicatrix with respect to a change in the ferroelectric polarization and the change of the ferroelectric polarization due to an electric field (the ferroelectric dielectric susceptibility). Importantly, this approach highlights the role that the linear susceptibilities in the ferroelectric polarization play in determining the electro-optic properties. [15] Further developments in the theory of optical properties led to a similar description for the piezo-optic effect. [16]

Microscopically, the polar-optic tensor is based upon the polarization potential description, [17] which relates the displacement of ions in the crystal, i.e., polarization, to the changes in the electronic band structure. Other microscopic models, such as the bond valance model, require the knowledge of the exact atomic positions and electronic wavefunctions [18,19]. Since the LGD theory only contains information about the ferroelectric polarization, the other microscopic approaches are poorly suited to describe the coupling between the electronic dielectric susceptibility and the ferroelectric polarization.

Despite the success of using the polar-optic tensor, it is not complete in its description of dispersion and cannot consistently include the effects of multiple stimuli. Dispersion plays a large role in the optical properties in the visible spectrum, particularly for perovskite ferroelectrics. Therefore, any complete description must account for the connection between the ferroelectric, optical, and dispersive properties. Currently, the dispersion of the polar-optic tensor relies upon a constant Sellmeier oscillator description, [14] which shows agreement with experimental measurements at room temperature. However, this approach assumes that the dispersion is constant with temperature, polarization, and stress, thus it is not applicable to cases with significant temperature variations, or across ferroelectric phase transitions. To fix this, one must make the dispersion dependent upon temperature, polarization, and stress. Here, a thermodynamic approach may resolve these issues to provide a more complete and comprehensive description of the optical properties.

Initial work by Garrett [20] made a simple 1-D thermodynamic model based upon the ionic and electronic polarizations as order parameters. This model provided a proof of concept for the idea of extending thermodynamics to the study of optical properties. The optical properties were calculated from the frequency-dependent linear and nonlinear dielectric susceptibility derived from classical and quantum models, both of which produced the same final expression. However, since this thermodynamic model is one-dimensional and largely neglected crystal symmetry, it cannot calculate the full nonlinear optical tensors or describe changes in optical properties due to phase transitions.

In this work, we present a comprehensive thermodynamic theory of optical properties of ferroelectrics. Our approach simultaneously incorporates electro-, thermo-, and piezo-optical properties, thereby simplifying and clarifying the connections between them, instead of treating each optical property in isolation as in existing theories. We separate the total polarization into the lattice and electronic contributions and connect the two with biquadratic coupling tensors. We validate the thermodynamic formulation by comparing results from first principles calculations and experimental measurements. In the thermodynamic limit, we derive the linear and electro-optic properties by taking the appropriate derivatives of the free energy function. From our analysis, we find that each contribution to the polarization corresponds to a particular effective mass, leading to dispersion in the polarization dynamics. Our thermodynamic formulation goes beyond previous methods by providing greater clarity on the connections between the ferroelectric, optical, and dispersive properties, allowing for the consistent and precise determination of the full temperature-dependent electro-optic tensor and other anisotropic optical properties. This thermodynamic free energy function may potentially be utilized in modeling optical properties to the future integration of ferroelectrics into photonic integrated circuits, where the polarization and strain states are significantly different from their corresponding bulk conditions.

## II. THEORETICAL MODEL
### A. Definition of Lattice and Electronic Polarizations

To develop a fundamental equation of thermodynamics, an important first step is to define the relevant order parameters. We separate the polarization into two components, $\vec{P} \equiv \vec{P}^L + \vec{P}^e$, where $\vec{P}^L$ is the polarization contribution that arises solely from the displacement of the crystal lattice and $\vec{P}^e$ corresponds to the field-induced displacements of the centers of charge of the electronic orbitals (the Wannier centers) from the zero-field equilibrium position.

We formally define the lattice and the electronic polarization through the lens of the modern theory of polarization. [21,22] The modern theory of polarization formally defines the total polarization as the sum of the Berry-phase moment from electrons and the point charge moment from the ion [23]

$$\vec{P}(\vec{E}) = \frac{e}{\Omega}\left(\sum_I Z_I \vec{R}_I(\vec{E}) - \sum_n \vec{r}_n(\{\vec{R}_I(\vec{E})\}, \vec{E})\right),$$ (2)

where $\vec{R}_I$ is the position of the $I^{\text{th}}$ ion as a function of the applied electric field, $\vec{r}_n$ is the center of the $n^{\text{th}}$ Wannier function which depends upon both the positions of the ions and the applied electric field, $Z_I$ is the valence of the $I^{\text{th}}$ ion, and $\Omega$ is the volume of the simulation cell. We define the lattice polarization as

$$\vec{P}^L(\vec{E}) = \frac{e}{\Omega}\left(\sum_I Z_I \vec{R}_I(\vec{E}) - \sum_n \vec{r}_n(\{\vec{R}_I(\vec{E})\}, \vec{0})\right),$$ (3)

which includes the change in polarization from the displacement of ions and the change in polarization due to the displacement of the Wannier center caused by the displacement of ions. From this definition, we assume that the electrons reach an equilibrium configuration instantaneously in response to the displacement of the ion. The induced electronic polarization is

$$\vec{P}^e(\vec{E}) = -\frac{e}{\Omega}\sum_n \left(\vec{r}_n(\{\vec{R}_I(\vec{E})\}, \vec{E}) - \vec{r}_n(\{\vec{R}_I(\vec{E})\}, \vec{0})\right).$$ (4)

The problem of insulators in an applied electric field is subtle, where under a uniform electric field, the system becomes non-periodic and unbounded from below. [23,24] The induced electronic polarization is not the true ground state as there is no ground state under a finite electric field; the energy may always be lowered by transferring charge from the valence band of one region to the conduction band states in another region. We follow the interpretation introduced by Nunes et al., [23] where the electronic polarization is generated by adiabatically applying the electric field while keeping the same periodicity of the system. For the thermodynamic (static) limit of the response, the field must oscillate slowly compared to typical electronic processes and fast compared to the rate of electronic tunneling.

Conceptually, for $\vec{P}^L$, when the ions in the crystal lattice are displaced, the distance between ions will change. The change in interatomic distance will shift the center of charge density in the crystal. Therefore $\vec{P}^L$ describes the ionic *and* electronic components of polarization that arise from the displacement of the crystal lattice alone. This lattice polarization is thus defined as a *hybrid mode* of ionic displacements and the corresponding electronic hybridizations driven by the ionic displacements, [25] which is also consistent with the definition of the Born effective charge, which relates the change in polarization to the atomic displacement. [26] The electronic polarization, $\vec{P}^e$, is the displacement of Wannier centers from their zero-field equilibrium position. This term may be understood as the induced electronic polarization.

From the definitions in equation 3 and 4, under zero electric field, $\vec{P}^e = 0$, $\vec{P} = \vec{P}^L = \vec{P}^\circ$ where $\vec{P}^\circ$ is the spontaneous polarization. In this case, at a fixed temperature and pressure, the magnitude of the spontaneous polarization is constant, but its direction may change as a function of an applied electric field. Under a finite electric field we find,

$$P_i^L = P_i^\circ + \chi_{ij}^{L,1} E_j + \chi_{ijk}^{L,2} E_j E_k + \cdots \tag{5}$$

$$P_i^e = \chi_{ij}^{e,1} E_j + \chi_{ijk}^{e,2} E_j E_k + \cdots, \tag{6}$$

where $\chi_{ij}^{e,1}$ and $\chi_{ij}^{L,1}$ are the linear dielectric susceptibility, and the higher-order terms represent the nonlinear dielectric susceptibility. To obtain the total dielectric displacement, we include the vacuum permittivity, which includes the contributions to the total dielectric displacement from neither the lattice nor the electrons. Thus

$$\vec{D} = \vec{P}^L + \vec{P}^e + \epsilon_0 \vec{E}. \tag{7}$$

This description, using the lattice and induced electronic polarizations, contrasts with previous models. The Wemple-DiDomenico model uses the spontaneous and induced polarization, defining the spontaneous polarization as the ferroelectric polarization in the absence of any external electric field and the induced polarization as the polarization induced from an electric field $\vec{P} = \vec{P}^\circ + \delta \vec{P}$. [14] This definition does not differentiate between the induced lattice and electronic polarizations, leaving it ill-suited to distinguish between optical properties occurring at different frequencies. First-principles calculations define the ionic and electronic polarization, [27] which can distinguish between optical properties occurring at different frequencies and is useful for theoretical calculations. However, this approach is challenging to interface with experiments where it is uncertain if the electronic and ionic polarization components can be measured separately, and when the overall behavior of the mode is incompatible with purely ionic motions. [25] The lattice and induced electronic polarization definition distinguishes between frequency-dependent optical phenomena and allows for each quantity to be experimentally determined, and interfaces with existing LGD free-energy functions.

**B. Thermodynamic Description of Optical Properties**

The real part of the refractive index determines the phase velocity with which electromagnetic waves propagate through a medium, while the imaginary part of the complex index leads to optical absorption. For simplicity, let us consider materials with a relative magnetic permeability equal to one, thus allowing the refractive index squared to equal to the relative dielectric susceptibility. Importantly, these definitions are made within the non-dispersive, no absorption limit, henceforth called the thermodynamic limit.

Considering a low-frequency electromagnetic wave ($\omega < \omega^L$), such that all contributions to the dielectric constant are active, we write,

$$(n^2)_{ij} = \frac{1}{\epsilon_0}\left(\frac{\partial D_i}{\partial E_j}\right) = \chi_{ij}^L + \chi_{ij}^e + \delta_{ij} \tag{8}$$

where $\delta_{ij}$ is the Kronecker delta, which is the contribution of the vacuum. Figure 2 provides a schematic diagram for the frequency dependent contributions to the total dielectric susceptibility.

At frequencies well above the lattice resonances and below the band gap($\omega^L < \omega < \omega^e$), the lattice polarization cannot respond, generally in the near infrared and visible, only the vacuum dielectric susceptibility and electronic susceptibility are observed, allowing the following relation:

$$(n^2)_{ij} = \frac{1}{\epsilon_0}\left(\frac{\partial D_i}{\partial E_j}\right)_{P^L} = \chi_{ij}^e + \delta_{ij} \qquad (9)$$

At high enough frequencies where the both the lattice and the electronic polarizations cannot respond, ($\omega^L \ll \omega^e \ll \omega$) such as non-resonant x-ray regime, we obtain

$$(n^2)_{ij} = \frac{1}{\epsilon_0}\left(\frac{\partial D_i}{\partial E_j}\right)_{P^L, P^e} = \delta_{ij} \qquad (10)$$

We note that above the band gap, there will be additional electronic transitions that are not accounted for in this description.

For all optical properties, it is critical to identify the optical frequency of interest. The majority of electro-optic devices are designed for light with frequencies in the near-infrared to visible spectrum. In this frequency range, the refractive index of interest will be primarily determined by the electronic dielectric susceptibility from Eq. 9. However since previous thermodynamic descriptions [5–7] did not separate the contributions from the lattice and electronic polarization, their refractive index is primarily dictated by the lattice dielectric susceptibility. Applying this assumption to calculate the refractive index, one would obtain a refractive index of around $10 - 30$, which is $5 - 15$ times the reported value for BaTiO$_3$ within the near-infrared to visible wavelengths. Since these methods cannot calculate the correct value for the refractive index, it is not reasonable to expect that the proper value for the electro-optic coefficient within the near-infrared to visible wavelengths (which describes the change in refractive index) can be calculated.

A more useful description is found through the optical indicatrix, or the relative dielectric stiffness is related to the inverse of the refractive index squared. In the near infrared and visible range, the optical indicatrix is

$$\left(B_{ij}\right)^{-1} = \left(B_{ij}^e\right)^{-1} + \delta_{ij} \qquad (11)$$

where we define the electronic dielectric stiffness tensor as the second derivative of the thermodynamic free energy density with respect to the electronic polarization.

$$B_{ij}^e = \epsilon_0\left(\frac{\partial E_i}{\partial P_j^e}\right) = \epsilon_0\left(\frac{\partial^2 f}{\partial P_i^e \partial P_j^e}\right) \qquad (12)$$

The changes in the lattice dielectric susceptibility of the lattice polarization are described by LGD theory, allowing for the prediction of the dielectric response under different conditions. We seek to extend the current LGD theory by considering the changes in the electronic dielectric stiffness. We relate the change in the electronic dielectric stiffness tensors to polarization through a Taylor expansion using the parent high symmetry phase as the reference state:

$$\Delta B_{ij}^e = B_{ij}^e - B_{ij}^{e,ref} = g_{ijkl}^{LL} P_l^L P_k^L + g_{ijkl}^{Le} P_l^L P_k^e + g_{ijkl}^{ee} P_k^e P_l^e + \pi_{ijkl}\sigma_{kl} + \dots \qquad (13)$$

The $g_{ijkl}$ tensors in equation 13 are defined as

$$g_{ijkl}^{LL} = \left(\frac{\partial^2 B_{ij}^e}{\partial P_k^L \partial P_l^L}\right)_{T,\sigma}, \quad g_{ijkl}^{Le} = \left(\frac{\partial^2 B_{ij}^e}{\partial P_k^L \partial P_l^e}\right)_{T,\sigma}, \quad g_{ijkl}^{ee} = \left(\frac{\partial^2 B_{ij}^e}{\partial P_k^e \partial P_l^e}\right)_{T,\sigma}, \tag{14}$$

where $\pi_{ijkl}$ is the elasto-optic tensor and $B_{ij}^{e,ref}$ is the electronic dielectric stiffness of the high symmetry parent phase. [14] This contrasts with the Wemple-DiDomenico theory which involves changes in the total dielectric stiffness in response to the lattice polarization (electronic contribution and vacuum permittivity), while we only consider the changes in the electronic dielectric stiffness. [14] From equation 9, if the electronic dielectric constant approaches zero, then the total dielectric constant must approach one, while in the Wemple-DiDomenico formalism, the total dielectric constant can approach zero, which is unphysical for non-resonant responses. This difference becomes important for the accurate description of phases with a large change in the dielectric stiffness or a large polarization, which typically occurs at low temperatures.

To account for thermal expansion effects, we write

$$B_{ij}^{e,ref}(T) = B_{ij}^{e,ref}(T_0) + p_{ijkl}\alpha_{kl}(T - T_0) \tag{15}$$

where $p_{ijkl}$ is the elasto-optic strain tensor of the high symmetry phase where $p_{ijkl} = \pi_{ijmn}C_{mnkl}$, $\alpha_{kl}$ is the thermal expansion tensor, and $T_0$ is a reference temperature. Now, we describe the total electronic dielectric stiffness as

$$B_{ij}^e = B_{ij}^{e,ref}(T) + g_{ijkl}^{LL}P_l^L P_k^L + g_{ijkl}^{Le}P_k^e P_l^L + g_{ijkl}^{ee}P_k^e P_l^e + \pi_{ijkl}\sigma_{kl} \tag{16}$$

Integrating the above equation and utilizing the Landau-Ginzburg-Devonshire theory to describe the stability of the lattice polarization, we develop a partial fundamental equation of thermodynamics, including both the lattice polarization and the electronic polarization.

$$
\begin{aligned}
f\left(T, P_i^L, P_j^e, \sigma_{ij}, E_i\right) = \\
= a_{ij}P_i^L P_j^L &+ a_{ijkl}P_i^L P_j^L P_k^L P_l^L + a_{ijklmn}P_i^L P_j^L P_k^L P_l^L P_m^L P_n^L + \dots \\
&+ \frac{1}{2\epsilon_0}\left(B_{ij}^{e,ref}(T) + g_{ijkl}^{LL}P_l^L P_k^L\right)P_i^e P_j^e + \frac{1}{6\epsilon_0}g_{ijkl}^{Le}P_i^e P_j^e P_k^e P_l^L + \frac{1}{24\epsilon_0}g_{ijkl}^{ee}P_i^e P_j^e P_k^e P_l^e \\
&- \frac{1}{2}s_{ijkl}\sigma_{ij}\sigma_{kl} - Q_{ijkl}\sigma_{ij}P_k^L P_l^L - \frac{1}{2\epsilon_0}\pi_{ijkl}\sigma_{ij}P_k^e P_l^e - E_i P_i^L - E_i P_i^e - \frac{1}{2}\epsilon_0\delta_{ij}E_i E_j
\end{aligned}
\tag{17}
$$

Using this fundamental equation of thermodynamics, one can account for other conditions, such as constant strain or mixed boundary (thin film) conditions, by using the appropriate Legendre transforms.

Density-functional theory (DFT) predictions for the polar-optic tensors of tetragonal BaTiO$_3$ are shown in Figure 3. Here, we implement a fully nonempirical inter-site Hubbard $U+V$ correction [28,29] to calculate the optical characteristics and dielectric function. First, we map the dielectric response along the minimum energy path connecting the cubic paraelectric phase to the ferroelectric tetragonal phase using the nudged elastic band method. Then, we calculate the refractive indices using the DFT+$U+V$ approach to analyze the relation between the electronic dielectric stiffness and the lattice polarization. The implementation of the self-consistent Hubbard $U+V$ correction enables the accurate prediction of the band structure without suppressing the orbital's hybridization, which captures the covalent nature of the Ti–O bond [29] and yields a more accurate estimation of the anisotropic refractive indices compared to standard DFT and DFT+$U$.

We calculate the electronic dielectric stiffness tensors with different lattice polarization values using density-functional perturbation theory to fit the polarization-dependent electronic dielectric stiffness. As shown in Figure 2. we find the behavior closely matches the parabolic dependence of the polarization up to a lattice polarization of 0.2 C/m², and beyond 0.2 C/m², an additional 6$^{\text{th}}$ rank polar-optic tensor may be required to fully describe the relationship.

We note the results of Veithen et al. concluded that the basic assumptions involved in the creation of the polar-optic tensor are not met in practice. [27] However, Veithen et al. used the optical dielectric susceptibility instead of the optical dielectric stiffness to derive the polar-optic tensor, which may significantly change the conclusions. Here, we find that using the quadratic relationship of the optical dielectric stiffness and polarization, the relationship originally described in the Wemple-DiDomenico model fits well with both the first principles results and the experimental data. The proposed free-energy function in our current work only considers the lowest order terms allowed by symmetry to describe the birefringence, second harmonic generation, and third harmonic generation. Higher order terms in the Taylor expansion, in principle, always exist, and future free energy functions may include these to more accurately describe the optical properties.

## C. Calculation of Optical Properties in the Thermodynamic Limit

Using the fundamental equation of thermodynamics established in Eq. 17, we may calculate the optical properties in the thermodynamic limit (non-dispersive and no absorption). At a given temperature and composition, the equilibrium lattice and electronic polarization are found by solving the following coupled equations.

$$\left(\frac{\partial f}{\partial P_i^L}\right)_{T,\sigma,P^e} = 0, \qquad \left(\frac{\partial f}{\partial P_i^e}\right)_{T,\sigma,P^L} = 0. \tag{18}$$

In general, the electro-optic effect can be split into two parts. First, an applied electric field will change the polarization, which is described by the dielectric susceptibility. Second, the polarization will change the optical dielectric response, which is described by the polar-optic effect.

The dielectric susceptibility is the inverse of the second derivative of the free energy density with respect to polarization. Separating between the lattice and electronic contributions, we find

$$\chi_{ij}^L = \frac{1}{\epsilon_0}\left(\frac{\partial P_i^L}{\partial E_j}\right)_{T,\sigma} = \frac{1}{\epsilon_0}\left(\frac{\partial^2 f}{\partial P_i^L \partial P_j^L}\right)_{T,\sigma}^{-1}, \tag{19}$$

$$\chi_{ij}^e = \frac{1}{\epsilon_0}\left(\frac{\partial P_i^e}{\partial E_j}\right)_{T,\sigma} = \frac{1}{\epsilon_0}\left(\frac{\partial^2 f}{\partial P_i^e \partial P_j^e}\right)_{T,\sigma}^{-1}. \tag{20}$$

We derive the index ellipsoid or optical dielectric stiffness, which is the primary interest in this work for applications and experimental measurements, by rearranging Eq. 11,

$$B_{ij} = (n^2)_{ij}^{-1} = B_{ik}^e[(B^e + \mathbb{I})^{-1}]_{kj}, \tag{21}$$

where $\mathbb{I}$ is the identity tensor. The polar-optic effect is the change of the index ellipsoid in response to a change in polarization. Here, we consider the polar optic effect from both the lattice polarization,

$$f_{ijk}^L = \left(\frac{\partial B_{ij}}{\partial P_k^L}\right)_{T,\sigma} = \left(\frac{\partial B_{ij}}{\partial B_{mn}^e}\right)_{T,\sigma} \left(\frac{\partial B_{mn}^e}{\partial P_k^L}\right)_{T,\sigma} = \epsilon_0 \left(\frac{\partial B_{ij}}{\partial B_{mn}^e}\right)_{T,\sigma} \left(\frac{\partial^2 f}{\partial P_m^e \partial P_n^e \partial P_k^L}\right)_{T,\sigma}$$

$$= [(B^e + \mathbb{I})^{-1}]_{mi}[(B^e + \mathbb{I})^{-1}]_{nj}[2g_{mnkl}^{LL}P_l^L + g_{mnkl}^{Le}P_l^e].$$

(22)

and the electronic polarization,

$$f_{ijk}^e = \left(\frac{\partial B_{ij}}{\partial P_k^e}\right)_{T,\sigma} = \left(\frac{\partial B_{ij}}{\partial B_{mn}^e}\right)_{T,\sigma} \left(\frac{\partial B_{mn}^e}{\partial P_k^e}\right)_{T,\sigma} = \epsilon_0 \left(\frac{\partial B_{ij}}{\partial B_{mn}^e}\right)_{T,\sigma} \left(\frac{\partial^2 f}{\partial P_m^e \partial P_n^e \partial P_k^e}\right)_{T,\sigma}$$

$$= [(B^e + \mathbb{I})^{-1}]_{mi}[(B^e + \mathbb{I})^{-1}]_{nj}[g_{mnkl}^{Le}P_l^L + 2g_{mnkl}^{ee}P_l^e].$$

(23)

Dropping the tensor notation for clarity, we rewrite Eq. 22 and 23

$$f^L = \frac{2g^{LL}P^L + g^{Le}P^e}{(B^e+1)^2}, \quad f^e = \frac{g^{Le}P^L + 2g^{ee}P^e}{(B^e+1)^2}.$$

(24)

Using the polar-optic effect and the dielectric susceptibility from each polarization component, we obtain the electro-optic effect

$$r_{ijk}^L = \left(\frac{\partial B_{ij}}{\partial P_m^L}\right)_{T,\sigma} \left(\frac{\partial P_m^L}{\partial E_k}\right)_{T,\sigma} = \varepsilon_0 f_{ijm}^L \chi_{mk}^L,$$

(25)

$$r_{ijk}^e = \left(\frac{\partial B_{ij}}{\partial P_m^e}\right)_{T,\sigma} \left(\frac{\partial P_m^e}{\partial E_k}\right)_{T,\sigma} = \varepsilon_0 f_{ijm}^e \chi_{mk}^e.$$

(26)

Therefore, the total electro-optic effect is

$$r_{ijk}^{total} = r_{ijk}^L + r_{ijk}^e$$

(27)

In ferroelectrics, $\chi_{mk}^e$ is typically on the order of $10^0$, and $\chi_{mk}^L$ is typically on the order of $10^3$-$10^4$ for ferroelectrics. Therefore, for modulation frequencies where the lattice polarization may respond, typically below the terahertz, one may approximate

$$r_{ijk}^{total} \cong r_{ijk}^L \ (\omega \leq \omega_L)$$

(28)

We note that the piezoelectric response is contained within the lattice response. One can use Legendre transforms to account for constant strain or mixed boundary (thin film) conditions depending upon the experimental conditions.

Other optical properties are also described in the thermodynamic free energy density function. For example, the thermal-optic effect is the change in the index ellipsoid due to a change in temperature. This is separated into thermal expansion based- and pyroelectric-responses

$$\left(\frac{\partial B_{ij}}{\partial T}\right)_{E,\sigma} = \epsilon_0 \left(\frac{\partial B_{ij}}{\partial B_{mn}^e}\right)_{E,\sigma} \left(\frac{\partial^2 f}{\partial P_m^e \partial P_n^e \partial T}\right)_{E,\sigma}$$

$$= p_{mnkl}\alpha_{kl}[(B^e + \mathbb{I})^{-1}]_{mi}[(B^e + \mathbb{I})^{-1}]_{nj} + f_{ijl}^L \phi_l,$$

(29)

where the pyroelectric effect is

$$\phi_i = \left(\frac{\partial^2 f}{\partial E_i \partial T}\right)_{E,\sigma} = \left(\frac{\partial P_i^L}{\partial T}\right)_{E,\sigma}. \tag{30}$$

Again, dropping the tensor notation for clarity, we find

$$\left(\frac{\partial B}{\partial T}\right) = \frac{p\alpha}{(B^e+1)^2} + f^L\phi. \tag{31}$$

The piezo-optic effect relates a change in the index ellipsoid to the change in stress, which may be separated into an intrinsic piezo-optic response and a response mediated by the piezoelectric effect. Using the thermodynamic free energy density function, we write

$$\begin{aligned}
\left(\frac{\partial B_{ij}}{\partial \sigma_{kl}}\right)_{T,E} &= \epsilon_0 \left(\frac{\partial B_{ij}}{\partial B_{mn}^e}\right)_{T,E} \left(\frac{\partial^3 f}{\partial P_m^e \partial P_n^e \partial \sigma_{kl}}\right)_{T,E} \\
&= \pi_{mnkl}[(B^e + \mathbb{I})^{-1}]_{mi}[(B^e + \mathbb{I})^{-1}]_{nj} + f_{ijm}^L d_{mkl}
\end{aligned} \tag{32}$$

where $d_{mkl}$ is the piezoelectric effect, which is equal to

$$d_{ijk} = -\left(\frac{\partial^2 f}{\partial E_i \partial \sigma_{jk}}\right)_{T,E} = \left(\frac{\partial P_i^L}{\partial \sigma_{jk}}\right)_{T,E}. \tag{33}$$

Dropping the tensor notation for clarity, we have

$$\left(\frac{\partial B}{\partial \sigma}\right) = \frac{\pi}{(B^e+1)^2} + f^L d. \tag{34}$$

The thermodynamic formulation of optical properties allows for a connection between the ferroelectric properties that are well described using existing thermodynamic free-energy density functions and the optical properties through the quadratic polar-optic tensor. In ferroelectrics, the change in optical properties is dominated by the change in ferroelectric polarization. We can decompose the change in optical properties under an external stimulus to the change in the optical properties due to the change in polarization and the change in ferroelectric properties due to the external stimuli, significantly simplifying and clarifying the calculation.

**D. Calculation of Frequency-Dependent Optical Properties**

To account for optical dispersion and loss, we extend our thermodynamic formulation using the polarization equation of motion, [30] which is analogous to the Lorentz model [31]

$$\mu_{ij}^e \frac{\partial^2 P_j^e}{\partial t^2} + \gamma_{ij}^e \frac{\partial P_j^e}{\partial t} = -\frac{\delta F}{\delta P_i^e}, \tag{35}$$

where $\mu_{ij}^e$ is the electronic polarization effective mass and $\gamma_{ij}^e$ is the electronic polarization damping constant. In the absence of a static electric field and at a sufficient frequency so that we assume the lattice polarization remains static, we obtain

$$\mu_{ij}^e \frac{\partial^2 P_j^e}{\partial t^2} + \gamma_{ij}^e \frac{\partial P_j^e}{\partial t} + \frac{1}{\epsilon_0} B_{ij}^e P_j^e + \frac{1}{\epsilon_0} g_{ijkl}^{Le} P_j^e P_k^e P_l^L + \frac{1}{\epsilon_0} g_{ijkl}^{ee} P_j^e P_k^e P_l^e = E_i. \tag{36}$$

Under sufficiently weak optical electric fields, the nonlinear terms will be much smaller than the linear term, allowing for the complex electronic dielectric susceptibility to be written as:

$$\widetilde{\chi_{ij}^{e,1}}(\omega) = \left[B_{ij}^e - \epsilon_0\left(i\omega\gamma_{ij}^e + \omega^2\mu_{ij}^e\right)\right]^{-1}.$$ (37)

Using the frequency-dependent electronic dielectric susceptibility, we find the frequency-dependent refractive index and electro-optic effect. Ignoring absorption, we obtain the polar-optic effect

$$f_{ijk}^L = \left(\frac{\partial B_{ij}(\omega)}{\partial P_k^L}\right)_{T,\sigma} = \left(\frac{\partial B_{ij}(\omega)}{\partial B_{mn}^e}\right)_{T,\sigma}\left(\frac{\partial B_{mn}^e}{\partial P_k^e}\right)_{T,\sigma}$$

$$= \left(\frac{\partial B_{ij}(\omega)}{\partial B_{mn}^e}\right)_{T,\sigma_{ij}}\left[2g_{mnkl}^{LL}P_l^L + g_{mnkl}^{Le}P_l^e\right].$$ (38)

where

$$\left(\frac{\partial B_{ij}(\omega)}{\partial B_{mn}^e}\right)_{T,\sigma} = [(B^e - \epsilon_0\omega^2\mu^e + \mathbb{I})^{-1}]_{mi}[(B^e - \epsilon_0\omega^2\mu^e + \mathbb{I})^{-1}]_{nj},$$ (39)

and the electro-optic effect,

$$r_{ijk}^L(\omega) = \left(\frac{\partial B_{ij}(\omega)}{\partial P_k^L}\right)_{T,\sigma}\left(\frac{\partial P_m^L}{\partial E_k}\right)_{T,\sigma} = f_{ijk}^L(\omega)\chi_{mk}^L.$$ (40)

Similar derivations can be made for the thermo-optic and piezo-optic tensors. The theory of Wemple and DiDomenico utilized a frequency-dependent polar optic tensor and demonstrated good agreement with the frequency-dependent electro-optic properties in experiments. [14] However, from equation 39, the dispersion in the electro-optic properties depends upon the electronic dielectric stiffness, which is dictated by the polarization, thus allowing for a consistent connection between the ferroelectric properties, optical properties, and dispersion. While a frequency dependent polar-optic tensor may fit well to experiments, especially at a single temperature, the accuracy is limited for its temperature dependence. In contrast, the thermodynamic approach presented here can be employed to calculate and capture the full temperature dependence of the dispersion and the interplay between the lattice polarization and the optical dispersion.

## III. RESULTS

BaTiO$_3$ has recently become a leading electro-optic material due to its large $r_{51}$ ($r_{131}$) electro-optic coefficient at room temperature. While the properties from room temperature to the paraelectric phase transition are well known, the optical properties of bulk BaTiO$_3$ below 270K are largely unknown. [3] Using the thermodynamic approach, we predict the full range of the temperature and wavelength-dependent linear optical and electro-optic properties. The expanded free energy density function and coefficients for BaTiO$_3$ are given in Appendix A.

The free energy density function coefficients are found by using the existing thermodynamic description of BaTiO$_3$ [32], and fitting the optical properties and dynamic coefficients to available experimental data [33–37]. Specifically, we fit our model from room temperature refractive indices and dispersion from [33–36,38], the electro-optic coefficients measured at 546.1 nm between 280K – 400K from [37], and the temperature dependent birefringence from [34,35]. We then predict the electro-optic coefficients at 633 nm [39], and the temperature dependent optical dispersion, and all optical properties below 280 K.

Using the thermodynamic description, we calculate the temperature-dependent ferroelectric and optical properties of BaTiO$_3$ shown in Figure 4. [32] Figure 4a shows the temperature-dependent lattice polarization with discontinuities at the ferroelectric phase transitions, which align with the change in the refractive indices with temperature shown in Figure 4b. Figure 4c shows the temperature-dependent birefringence at various wavelengths and shows excellent agreement with the experimental values at 633nm, demonstrating an increased birefringence at shorter wavelengths. Given that the current model includes only the lowest order terms, the value of the refractive indices at low temperatures may deviate from experiments since according to the DFT calculations, the relation between the polarization and optical properties may go beyond the quadratic regime. Future work may further investigate the low-temperature regions to fit the higher-order polar-optic tensors.

Figure 5 calculates the temperature-dependent lattice dielectric constant and the electro-optic effect. Near the tetragonal to cubic phase transition, we find the $\chi_{33}^L$ is enhanced (Fig. 5a), leading to a large $r_c$ electro-optic coefficient. As temperature decreases, approaching the tetragonal to orthorhombic, we find the dielectric constant $\chi_{11}^L$ is significantly enhanced, which drives the enhancement of the $r_{51}$ coefficient (Fig. 5b). In the orthorhombic phase, approaching the orthorhombic to rhombohedral phase transition, $\chi_{11}^L$ decreases while $\chi_{22}^L$ increases, leading to a decreased $r_{51}$ coefficient while $r_{42}$ increases with approaching a maximum of $r_{42}$=2485 pm/V near 200K. In the rhombohedral phase, the electro-optic coefficient is significantly diminished, and at cryogenic temperatures (4K), we predict the electro-optic coefficient $r_{51}$=125 pm/V, showing a close agreement to the effective electro-optic coefficient reported in BaTiO₃ thin films reported by Eltes et al. of $r_{eff}$=200 pm/V. [3] Overall, we demonstrate a close agreement to the values reported by Johnston at 546.1 nm, [38] and further agreement to experiments is demonstrated in Table1 comparing between the thermodynamic approach and experiments at 633 nm. [39] Although the use of only a quadratic polar-optic tensor may cause deviations in the refractive indices when the polarization becomes large, the temperature-dependent electro-optic properties are driven by the large temperature-dependent changes in the lattice dielectric constant. Therefore, we expect the electro-optic properties to agree with experiments at low temperatures.

**Table 1. Electro-optic Coefficients of BaTiO₃ from this work and experiments at 633nm**

|          | This work  | Experiment [39]     |
| -------- | ---------- | ------------------- |
| $r_{13}$ | 15 pm/V    | 8±2 pm/V            |
| $r_{33}$ | 109 pm/V   | 105 ± 10 pm/V       |
| $r_{51}$ | 1396 pm/V  | 1300 ± 100 pm/V     |

Through using the high symmetry cubic phase to determine the allowed tensor coefficients and the lattice polarization as a symmetry-breaking distortion, our description automatically accounts for all symmetry-allowed coefficients which is highlighted in Appendix D.

To further explore the validity of the electronic polarization dynamic equation, we directly compare it to the experimental dispersion. Figure 6a compares the experimental dispersion curves to the dispersion calculated from Eq. 37, showing an excellent agreement with the collected experimental data and a close agreement with the curves computed from DFT, demonstrating the alignment between the thermodynamic- and DFT-based calculations. We highlight the values calculated for the thermodynamic limit with the dashed lines, which approach the experimental values at wavelengths above 1500nm. Figure 6b shows the temperature-dependent dispersion in the extraordinary refractive index (along the tetragonal $c$-axis) calculated from Eq. 37 and Figure 6c shows the extraordinary refractive index measured from experiment. The calculated and experimental refractive indices are in close agreement showing an increase in the extraordinary refractive index and resonance wavelength with temperature, demonstrating a close connection between the ferroelectric polarization, optical properties, and dispersion.

## IV. DISCUSSION

The current thermodynamic approach to the electro-optic effect is conceptually different from existing thermodynamic approaches based only on LGD theory in terms of the lattice polarization. For example, previous thermodynamic models concluded that the electro-optic response is dominated by anharmonicity in the lattice polarization energy landscape. This is only true for light wavelengths larger than 10 μm (or frequencies less than 30 THz) where the lattice polarization dominates the total dielectric response. [5–7] However, for light frequencies in the near-infrared and visible spectrum, well above the resonances due to optical phonons, the electro-optic response is dominated by a biquadratic coupling between the lattice polarization and the electronic dielectric susceptibility. Therefore, the electro-optic response is dominated by the change in the lattice polarization in response to a low frequency modulating field (the linear dielectric susceptibility of the lattice polarization). In contrast to previous thermodynamic approaches, we find the anharmonicity in the lattice polarization energy landscape does *not* significantly contribute to the total electro-optic response.

We must also note the relationship between the dielectric susceptibility of the lattice and the maximum modulating frequency. Thus far, we have calculated the electro-optic response for a static modulating field, but applications in photonic integrated circuits (PICs) may require frequency modulation in the GHz frequency range. From the polarization dynamic equation, one may find a tradeoff between the dielectric constant and the maximum response frequency. Therefore, one would expect a similar tradeoff between the magnitude of the electro-optic effect and maximum modulating frequency. One future direction is to employ the dynamical phase field method to further investigate the relationship between the magnitude of the electro-optic effect and the maximum modulating frequency and the impact of the ferroelectric domain structure which may lead to emergent electro-optic behavior. [40]

## V. CONCLUSION

We formulated a new thermodynamic theory for the optical properties of ferroelectric crystals by separating the contributions of the lattice and electronic polarization. We highlight the importance of the polar optic effect in determining the elastic, electric, and thermal responses of ferroelectric crystals. The dynamics and resonance behavior of the electronic polarization are accurately described by the thermodynamic energy landscape, the effective masses, and damping coefficients, connecting the optical properties in the thermodynamic limit and their associated dispersion behavior. This thermodynamic framework will help guide future research to optimize the optical properties of ferroelectric materials. Furthermore, it will serve as the basis for modeling the optical properties using the phase-field method, enabling the study of optical processes in inhomogeneous crystals containing defects, [41] multidomain/multigrain structures by taking into account of ferroelectric polarization dynamics [40] as well as free carrier generation and dynamics [42]. Importantly, this framework is not limited to ferroelectric materials or purely linear optical properties. For example, the thermodynamic framework can be generalized to analyze the optical properties of non-ferroelectric nonlinear materials such as quartz or beta-barium Borate (BBO) by assigning an order parameter to describe the nature of the symmetry-breaking distortions of a reference high symmetry phase [43,44]. We may build upon insights gained from previous explorations in $Zn_{1-x}Mg_xO$, which resulted in simultaneous enhancements in nonlinear optical properties and piezoelectricity, to further design and optimize multifunctional optical platforms [45]. The general thermodynamic framework for predicting linear and nonlinear optical properties of ferroelectrics developed in this work can be one of the important future tools for discovering new materials with tailored and optimized optical properties.

**Acknowledgment**

The work was primarily supported as part of the Computational Materials Sciences Program funded by the U.S. Department of Energy, Office of Science, Basic Energy Sciences, under Award DE-SC0020145 (A.R., M.S.A., A.S., R.Z., V.G., I.D., and L.Q.C). A.R. is also partially supported by NSF under grant number DMR-2133373 and acknowledges the support of the National Science Foundation Graduate Research Fellowship Program under Grant No. DGE1255832.

## Appendix A: Detailed expression of the free energy function and the associated parameters

For BaTiO$_3$, we use the cubic phase as our high symmetry reference state and employ an 8th-order landau expansion [32] to describe the relative stability of the lattice polarization compared to the cubic reference state. For this manuscript, we only consider the lowest-order polar-optic tensor, which is sufficient to describe the linear optical and electro-optic properties. Based upon the symmetry of the reference phase we may expand equation 17 as

$$
\begin{aligned}
f\left(T, P_i^L P_i^e, E_i, \sigma_{ij}\right) = {} & f_o + a_{11}(T)[(P_1^L)^2 + (P_2^L)^2 + (P_3^L)^2] + a_{1111}[(P_1^L)^4 + (P_2^L)^4 + (P_3^L)^4] \\
& + a_{1122}[(P_1^L)^2(P_2^L)^2 + (P_2^L)^2(P_3^L)^2 + (P_1^L)^2(P_3^L)^2] + a_{111111}[(P_1^L)^6 + (P_2^L)^6 + (P_2^L)^6] \\
& + a_{111122}[(P_1^L)^4((P_2^L)^2 + (P_3^L)^2) + (P_2^L)^4((P_1^L)^2 + (P_3^L)^2) + (P_3^L)^4((P_1^L)^2 + (P_2^L)^2)] \\
& + a_{112233}[(P_1^L)^2(P_2^L)^2(P_3^L)^2] + a_{11111111}[(P_1^L)^8 + (P_2^L)^8 + (P_2^L)^8] \\
& + a_{11111122}[(P_1^L)^6((P_2^L)^2 + (P_3^L)^2) + (P_2^L)^6((P_1^L)^2 + (P_3^L)^2) + (P_3^L)^6((P_1^L)^2 + (P_2^L)^2)] \\
& + a_{11112222}[(P_1^L)^4(P_2^L)^4 + (P_2^L)^4(P_3^L)^4 + (P_1^L)^4(P_3^L)^4] \\
& + a_{11112233}[(P_1^L)^4(P_2^L)^2(P_3^L)^2 + (P_2^L)^4(P_1^L)^2(P_3^L)^2 + (P_3^L)^4(P_1^L)^2(P_2^L)^2] \\
& + \frac{1}{2\epsilon_o}B_{11}^{ref}(T)[(P_1^e)^2 + (P_2^e)^2 + (P_2^e)^2] + \frac{1}{2\epsilon_o}g_{1111}^{LL}[(P_1^L)^2(P_1^e)^2 + (P_2^L)^2(P_2^e)^2 + \\
& (P_2^L)^2(P_3^e)^2] \\
& + \frac{1}{2\epsilon_o}g_{1122}^{LL}[(P_1^L)^2((P_2^e)^2 + (P_3^e)^2) + (P_2^L)^2((P_1^e)^2 + (P_3^e)^2) + (P_3^L)^2((P_1^e)^2 + (P_2^e)^2)] \\
& + \frac{2}{\epsilon_o}g_{1212}^{LL}[P_1^L P_2^L P_1^e P_2^e + P_2^L P_3^L P_2^e P_3^e + P_1^L P_3^L P_1^e P_3^e] \\
& - \frac{1}{2}s_{1111}(\sigma_{11}^2 + \sigma_{22}^2 + \sigma_{33}^2) - s_{1122}(\sigma_{11}\sigma_{22} + \sigma_{22}\sigma_{33} + \sigma_{11}\sigma_{33}) - 2s_{1212}(\sigma_{12}^2 + \sigma_{23}^2 + \sigma_{13}^2) \\
& - Q_{1111}[\sigma_{11}(P_1^L)^2 + \sigma_{22}(P_2^L)^2 + \sigma_{33}(P_3^L)^2] \\
& - Q_{1122}[\sigma_{11}((P_2^L)^2 + (P_3^L)^2) + \sigma_{22}((P_1^L)^2 + (P_3^L)^2) + \sigma_{33}((P_1^L)^2 + (P_2^L)^2)] \\
& - Q_{1122}[\sigma_{11}((P_2^L)^2 + (P_3^L)^2) + \sigma_{22}((P_1^L)^2 + (P_3^L)^2) + \sigma_{33}((P_1^L)^2 + (P_2^L)^2)] \\
& - 2Q_{1212}[\sigma_{12}P_1^L P_2^L + \sigma_{23}P_2^L P_3^L + \sigma_{13}P_1^L P_3^L] \\
& - \frac{1}{2\epsilon_o}\pi_{1111}[\sigma_{11}(P_1^e)^2 + \sigma_{22}(P_2^e)^2 + \sigma_{33}(P_3^e)^2] \\
& - \frac{1}{2\epsilon_o}\pi_{1122}[\sigma_{11}((P_2^e)^2 + (P_3^e)^2) + \sigma_{22}((P_1^e)^2 + (P_3^e)^2) + \sigma_{33}((P_1^e)^2 + (P_2^e)^2)] \\
& - \frac{1}{\epsilon_o}\pi_{1212}[\sigma_{12}P_1^e P_2^e + \sigma_{23}P_2^e P_3^e + \sigma_{13}P_1^e P_3^e] \\
& - E_1 P_1^L - E_2 P_2^L - E_3 P_3^L \\
& - E_1 P_1^e - E_2 P_2^e - E_3 P_3^e \\
& - \epsilon_0(E_1^2 + E_2^2 + E_3^2) \,.
\end{aligned}
\tag{A1}
$$

where the coefficients used for this paper are given in Table A1.

**Table A1.** Coefficients in the thermodynamic free energy function and equation of motion

| | | | |
|---|---|---|---|
| $g_{1111}^{LL}$ | $18.5 \times 10^{-2} (m^4/C^2)$ | $p_{1111}$ | 0.5328 (Unitless) |
| $g_{1122}^{LL}$ | $2.5 \times 10^{-2} (m^4/C^2)$ | $p_{1122}$ | 0.1584 (Unitless) |
| $g_{1212}^{LL}$ | $12.85 \times 10^{-2} (m^4/C^2)$ | $p_{1212}$ | $-0.432$ (Unitless) |
| $\mu_e$ | $35.5 \times 10^{-23} \left(\frac{Kg\,m^4}{m\,C^2}\right)$ | $\gamma_e$ | $3 \times 10^{-9} \left(\frac{Kg\,m^4}{ms\,C^2}\right)$ |
| $a_{11}$ | $a_0(T - T_c)$ | $B_{ij}^{e,\,ref}(T_0)$ | 0.2356 (Unitless) |

| $T_c$ | 388$K$ | $T_0$ | 398$K$ |
|---|---|---|---|
| $a_0$ | $4.124 \times 10^5 \left(\frac{J}{m^3} \frac{m^4}{K\,C^2}\right)$ | $\alpha_{11}$ | $2.657 \times 10^{-5} \left(\frac{1}{K}\right)$ |
| $a_{1111}$ | $-2.097 \times 10^8 \left(\frac{J}{m^3} \frac{m^8}{C^4}\right)$ | $a_{11111111}$ | $3.863 \times 10^{10} \left(\frac{J}{m^3} \frac{m^{16}}{C^8}\right)$ |
| $a_{1122}$ | $7.974 \times 10^8 \left(\frac{J}{m^3} \frac{m^8}{C^4}\right)$ | $a_{11111122}$ | $2.529 \times 10^{10} \left(\frac{J}{m^3} \frac{m^{16}}{C^8}\right)$ |
| $a_{111111}$ | $1.294 \times 10^9 \left(\frac{J}{m^3} \frac{m^{12}}{C^6}\right)$ | $a_{11112222}$ | $1.637 \times 10^{10} \left(\frac{J}{m^3} \frac{m^{16}}{C^8}\right)$ |
| $a_{111122}$ | $-1.95 \times 10^9 \left(\frac{J}{m^3} \frac{m^{12}}{C^6}\right)$ | $a_{11112222}$ | $1.637 \times 10^{10} \left(\frac{J}{m^3} \frac{m^{16}}{C^8}\right)$ |
| $a_{112233}$ | $-2.509 \times 10^9 \left(\frac{J}{m^3} \frac{m^{12}}{C^6}\right)$ | $a_{11112233}$ | $1.367 \times 10^{10} \left(\frac{J}{m^3} \frac{m^{16}}{C^8}\right)$ |

**Appendix B: DFT Calculation Details**

DFT calculations were performed using QUANTUM ESPRESSO. [46] The plane-wave pseudopotential method is implemented, where we used the generalized-gradient approximation for the exchange-correlation functional, with the Perdew-Burke-Ernzerhof parametrization revised for solids (PBEsol) and ultrasoft pseudopotentials from the SSSP PRECISION library (version 1.1.2). [47–49] Kinetic energy and charge density cutoffs of 80 and 960 Ry, respectively, were selected. The Brillouin-zone sampling for computing electronic ground states used a Monkhorst-Pack centered 8x8x8 k-point mesh. After performing full geometry optimization at the DFT level of theory, the on-site and intersite Hubbard-parameter ($U + V$) calculations were completed using DFPT as implemented in the HP code [50], which is part of QUANTUM ESPRESSO. The dielectric response in the momentum space is calculated using the independent-particle approximation as a function of the frequency $\boldsymbol{\omega}$.

$$\boldsymbol{\epsilon}_{ij}(\boldsymbol{\omega}) = \epsilon_0 \delta_{ij} - \frac{e^2 \hbar^2}{\Omega m_e^2} \sum_{n,m} \frac{1}{(\varepsilon_n - \varepsilon_m)^2} \left( \frac{\langle \varphi_m | \boldsymbol{P}_i | \varphi_n \rangle \langle \varphi_n | \boldsymbol{P}_j | \varphi_m \rangle}{\varepsilon_n - \varepsilon_m + \hbar\omega + i\hbar\Gamma} + \frac{\langle \varphi_n | \boldsymbol{P}_i | \varphi_m \rangle \langle \varphi_m | \boldsymbol{P}_j | \varphi_n \rangle}{\varepsilon_n - \varepsilon_m - \hbar\omega - i\hbar\Gamma} \right) \quad \text{(B1)}$$

where the double index reveals the tensorial nature of $\boldsymbol{\epsilon}_{ij}(\boldsymbol{\omega})$, $\boldsymbol{P}_i$ and $\boldsymbol{P}_j$ are the polarizations in $\boldsymbol{\alpha}$ and $\boldsymbol{\beta}$ directions respectively, and $\Gamma$ is the damping parameter. The dielectric response is expressed in terms of its real and imaginary parts as:

$$\boldsymbol{\epsilon}(\boldsymbol{\omega}) = \boldsymbol{\epsilon_1}(\boldsymbol{\omega}) + i\,\boldsymbol{\epsilon_2}(\boldsymbol{\omega}) \quad \text{(B2)}$$

Consequently, the refractive index $\boldsymbol{n}(\omega)$ is calculated as

$$\boldsymbol{n}(\boldsymbol{\omega}) = \frac{1}{\sqrt{2}} \left( \frac{1}{\epsilon_0} \left[ \sqrt{\boldsymbol{\epsilon_1}(\boldsymbol{\omega})^2 + \boldsymbol{\epsilon_2}(\boldsymbol{\omega})^2} + \boldsymbol{\epsilon_1}(\boldsymbol{\omega}) \right] \right)^{1/2} \quad \text{(B3)}$$

**Appendix C: Experimental Details:**

**Sample preparation and SHG characterization**

A BaTiO$_3$ single crystal oriented along (100) out-of-plane was obtained from MTI Corporation. The crystal was grown using the top-seeded solution growth method and is single-side polished to minimize backside reflection. The crystal was then electrically poled by heating it to 150°C (T$_c$ = 130°C) and applying an in-plane electric field of 60 kV/m. The field was retained as the crystal was cooled to room temperature. Second harmonic generation (SHG) polarimetry was used to confirm the in-plane orientation of the polar axis (c-axis) at room temperature after poling.

The SHG polarimetry was measured using an 800 nm fundamental beam from a Ti:sapphire laser system (80 MHz, 100 fs). The linearly polarized incident beam was rotated using a half waveplate and focused on the sample surface. The sample was oriented such that the poling direction was parallel to the plane of incidence and made a 45° angle with the incident beam (supplementary figure 1a shows a schematic of the experimental geometry) and the second harmonic signal generated by the sample at 45° incidence was collected using a photo-multiplier tube. An analyzer positioned before the detector allows for the selection of s-polarized and p-polarized SHG intensities. The experimentally obtained polarimetry curves were compared to simulated curves generated by the #SHAARP open source package for SHG polarimetry analysis [51,52] for various optic axis (c-axis) orientations. The SHG coefficients for BaTiO$_3$ were taken from Miller [53] and the Herman and Hayden assumption was used for the calculations to account for the optical anisotropy of BaTiO$_3$ [54]. Supplementary figure 1b shows the simulated SHG polarimetry curves for the experimental conditions described above for different sample orientations. Good agreement between experimental data and simulations is obtained for the case where the polar axis is parallel to the plane of incidence and is 45° with respect to the incoming beam. This confirms the direction of the optic axis of the poled crystal.

**Temperature-dependent refractive indices**

The temperature-dependent optical constants of BaTiO$_3$ were measured using spectroscopic ellipsometry. The measurements were performed using a rotating compensator ellipsometer (J.A.Woollam M-2000) over the spectral range from 193 nm to 1000 nm at a single incidence angle of 64.4°. The sample temperature was controlled using a THMS600PS heating stage from Linkam Scientific. At each temperature measured, the general ellipsometry spectra were recorded at four points on the sample and at two different optic axis orientations – one with the c-axis lying within the plane of incidence (Euler angles θ=90°, ϕ=90°) and the other with the c-axis perpendicular to the plane of incidence (θ=90°, ϕ=0°). The two different optic axis orientations were accessed by rotating the crystal as the c-axis lies within the sample plane for both configurations.

The spectroscopic ellipsometry data was analyzed using J.A.Woollam CompleteEASE software package by constructing a model consisting of a semi-infinite uniaxial substrate. The ordinary and extraordinary optical constants were modeled using a Lorentz oscillator (Eq. 37), assuming only the stiffness $B_o$ and $B_e$ to be temperature dependent while constraining $m_o = m_e = m$ and $\gamma_o = \gamma_e = \gamma$, both $m$ and $\gamma$ being temperature independent. These values are determined by simultaneously fitting the general ellipsometry spectra at both the optic axis orientations and all temperatures using a least squares minimization routine for each of the four points independently. The fitting was restricted to above 500 nm to neglect absorption features. A comparison between the refractive indices at 35°C calculated using a Lorentz oscillator (as described above) and a Sellmeier model (as used in previous studies [34,35,38]) is shown in supplementary figure 2. The oscillator parameters for the multiple sample locations were averaged using an averaging scheme based on the reciprocal of the fitting errors squared as weights and the optical constants were determined from Eq. 37 using the weighted average values. The error bars for the refractive indices are calculated by taking the total differential of $\tilde{\chi}$ and using the weighted standard deviations of the fit parameters as the uncertainties.

**Appendix D: Linear Optical and Electro-optic Tensors**

**Table D1: Linear Optical and Electro-optic Tensors of the Tetragonal, Orthorhombic, and Rhombohedral Phases**

| Phase of BaTiO$_3$ | Linear Optical Dielectric Tensor $\lambda = 633$nm | Electro-optic tensor (pm/V) $\lambda = 633$nm |
|---|---|---|
| Cubic $Pm3m$ $P_i^L = (0,0,0)$ $T$=400K | $\begin{pmatrix} 5.813 & 0 & 0 \\ 0 & 5.813 & 0 \\ 0 & 0 & 5.813 \end{pmatrix}$ | $\begin{pmatrix} 0 & 0 & 0 \\ 0 & 0 & 0 \\ 0 & 0 & 0 \\ 0 & 0 & 0 \\ 0 & 0 & 0 \\ 0 & 0 & 0 \end{pmatrix}$ |
| Tetragonal $P4mm$ $P_i^L = (0,0,P)$ $T$=300K | $\begin{pmatrix} 5.791 & 0 & 0 \\ 0 & 5.791 & 0 \\ 0 & 0 & 5.556 \end{pmatrix}$ | $\begin{pmatrix} 0 & 0 & 15.0 \\ 0 & 0 & 15.0 \\ 0 & 0 & 109.1 \\ 0 & 1396.6 & 0 \\ 1396.6 & 0 & 0 \\ 0 & 0 & 0 \end{pmatrix}$ |
| Orthorhombic $Amm2$ $P_i^L = (P,0,P)$ $T$=240K | $\begin{pmatrix} 5.311 & 0 & 0 \\ 0 & 5.901 & 0 \\ 0 & 0 & 5.778 \end{pmatrix}$ | $\begin{pmatrix} 0 & 0 & -18.0 \\ 0 & 0 & 19.0 \\ 0 & 0 & 171.5 \\ 0 & 1479.6 & 0 \\ 237.2 & 0 & 0 \\ 0 & 0 & 0 \end{pmatrix}$ |
| Rhombohedral $R3m$ $P_i^L = (P,P,P)$ $T$=80K | $\begin{pmatrix} 5.900 & 0 & 0 \\ 0 & 5.900 & 0 \\ 0 & 0 & 5.018 \end{pmatrix}$ | $\begin{pmatrix} 0 & -42.9 & -3.7 \\ 0 & 42.9 & -3.7 \\ 0 & 0 & 119.6 \\ 0 & 174.1 & 0 \\ 174.1 & 0 & 0 \\ -42.9 & 0 & 0 \end{pmatrix}$ |

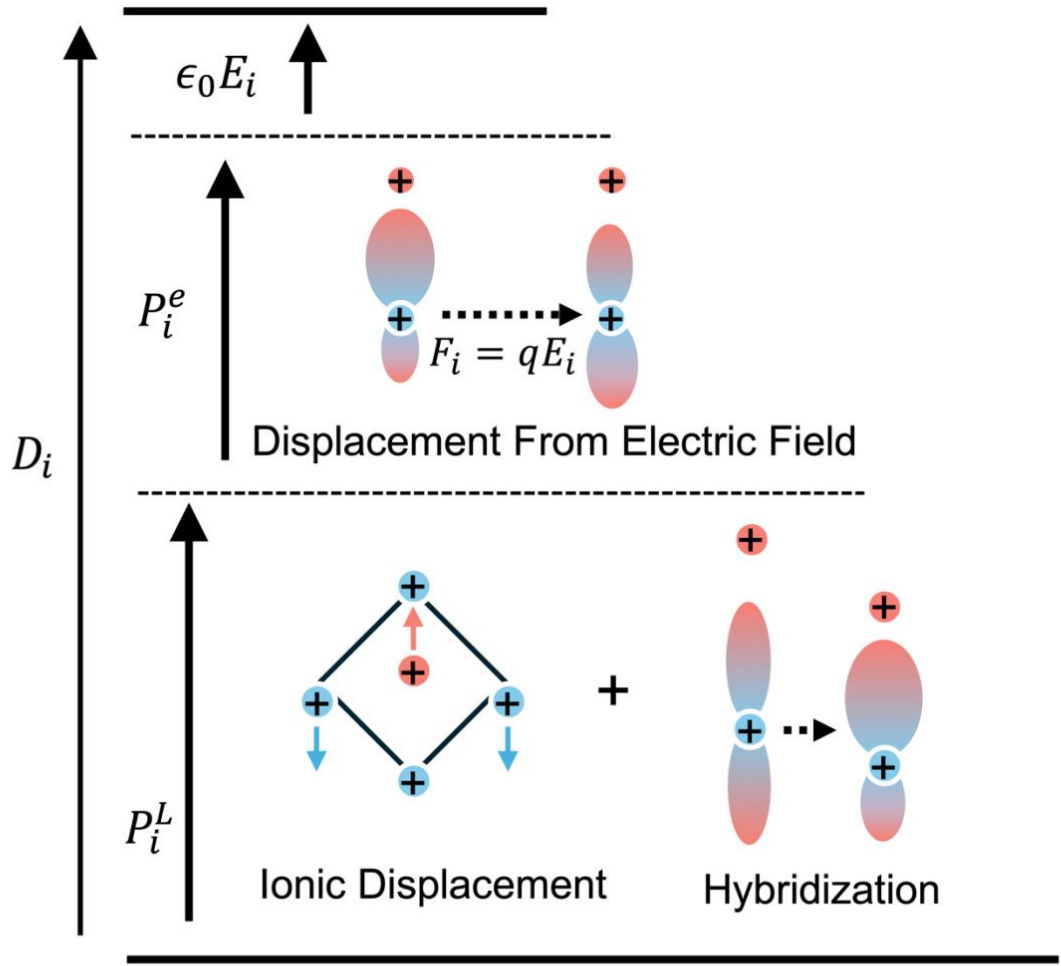

**Figure 1.** Contributions to the total dielectric displacement. Bottom: Lattice polarization. Middle: induced electronic polarization. Top: contribution from the vacuum

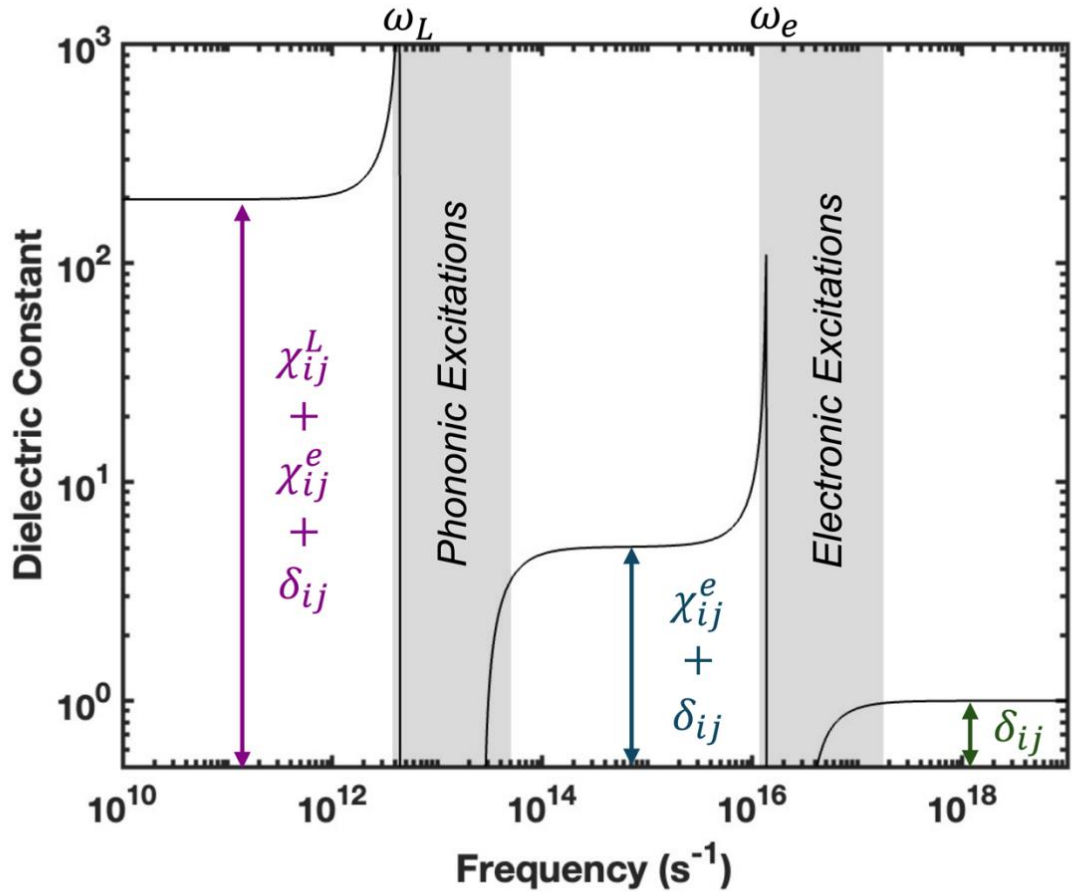

**Figure 2. Schematic Diagram of the frequency dependent contributions to the total dielectric constant.** Black curve denotes the frequency dependent contributions from the current model. $\omega_L$ is the resonance frequency of the lattice polarization and $\omega_e$ is the resonance frequency of the electronic polarization. Gray areas denote frequency regimes which tend to have additional excitations not accounted for in the model.

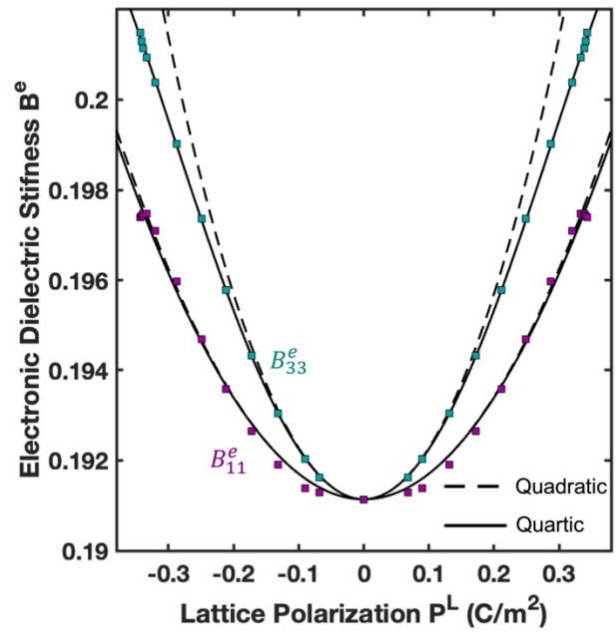

**Figure 3.** Dependence of the electronic dielectric stiffness on the lattice polarization for tetragonal BaTiO$_3$ using the DFT+U+V approach. The dashed line corresponds to the 2$^{nd}$-rank polar optic tensor, and the solid line corresponds to a 6$^{th}$-rank polar optic tensor description.

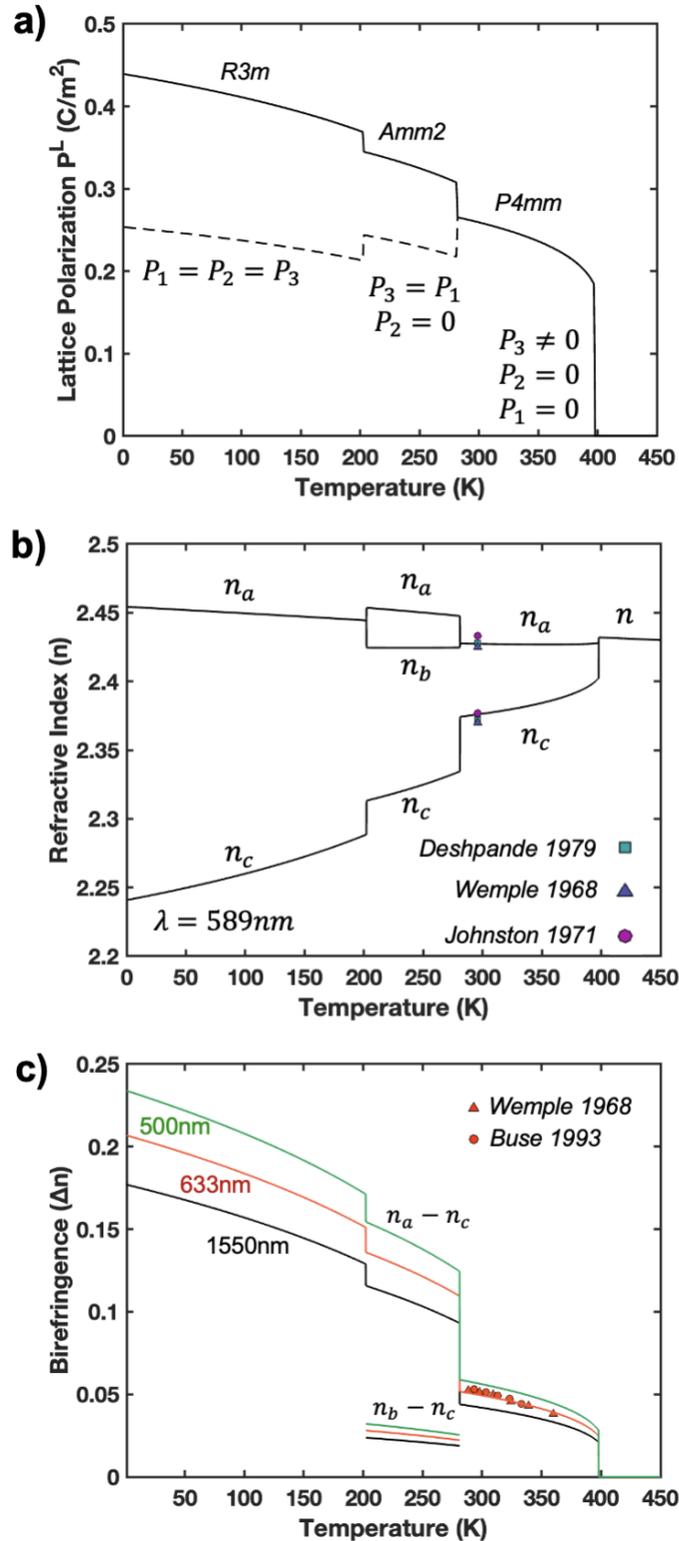

**Figure 4. Temperature-dependent polarization and refractive indices of BaTiO₃** a) lattice polarization, b) refractive index at 589nm with experimental data from [34,36,38] c) Temperature-dependent birefringence at various wavelengths with experimental data from [34,35]

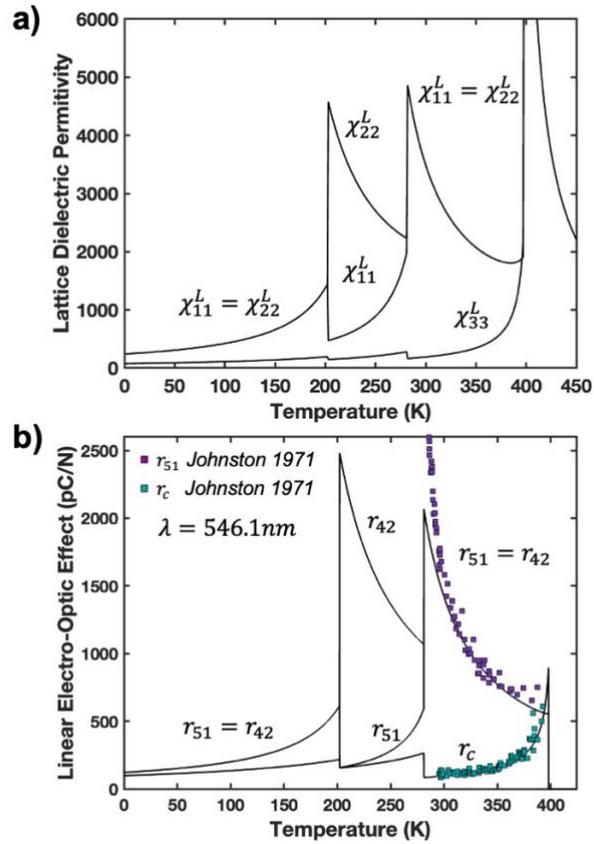

**Figure 5. Temperature-dependent dielectric and electro-optic properties of BaTiO₃** a) lattice dielectric constant, b) Temperature electro-optic effect at $\lambda =546.1$ nm ($r_c = r_{333} - r_{113}$) with experimental data from [37]

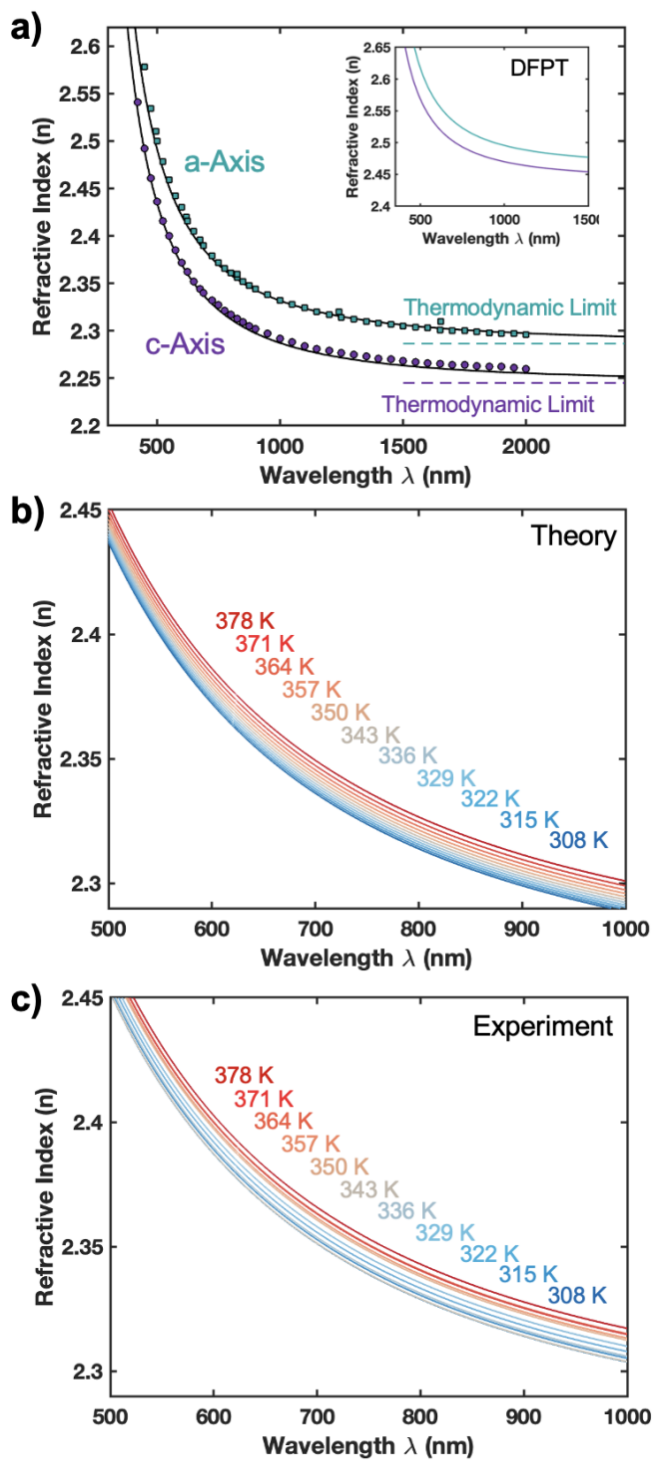

**Figure 6. Wavelength and temperature-dependent refractive indices** a) dispersion of the refractive index along the a-axis and c-axis. Solid lines correspond to the calculated dispersion, squares correspond to experimental data from [33], and dashed lines correspond to the thermodynamic limit. The inset shows the calculated dispersion from DFPT. b) temperature and frequency dependent dielectric dispersion of the extraordinary refractive index ($n_c$) from theory and c) from experiment

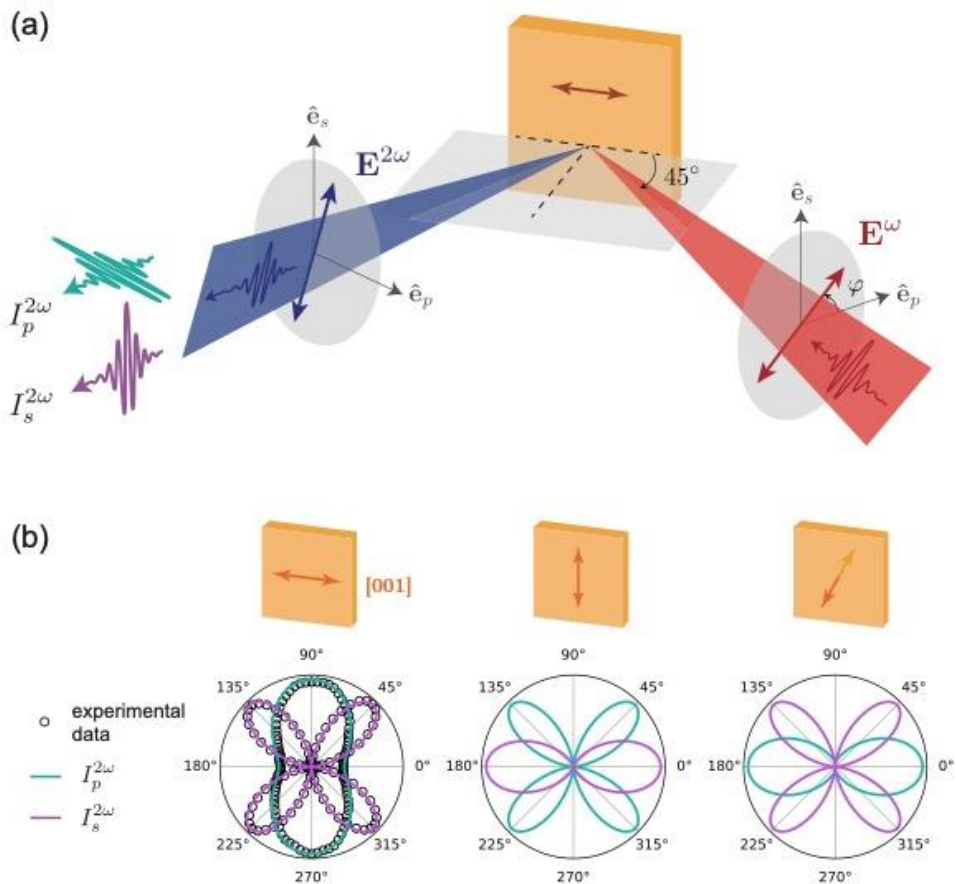

**Appendix figure 1: Second harmonic generation (SHG) measurement**
(a) Schematic of the optical SHG setup used to confirm the orientation of the optic axis. The (in-plane) poling direction is 45° with respect to the incident beam and is parallel to the plane of incidence. φ is defined as angle the incident electric field $\mathbf{E}^{\omega}$ makes with a unit vector $\hat{e}_p$ on the plane of incidence and orthogonal to the incident wavevector. (b) *s* and *p* polarized SHG intensities obtained from #SHAARP plotted as a function of the incident electric field rotation φ for common optic axis orientations observed in BaTiO₃. Agreement with the experimental data (black circles) is only seen for the case where the optic axis lies within the sample plane and the plane of incidence. All intensities are normalized to their respective maximum values.

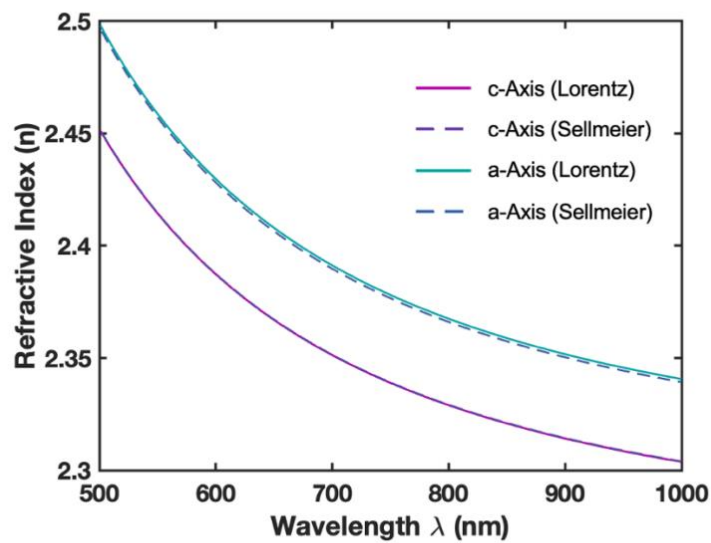

**Appendix figure 2: Comparison between refractive indices at 35°C determined using a Lorentz oscillator and a single oscillator Sellmeier model.**
To obtain the Sellmeier parameters, a method similar to the Lorentz oscillator case is used where only the ordinary and extraordinary amplitudes are allowed to vary as a function of temperature while the central wavelengths are constrained to a constant value determined by simultaneously fitting the spectra at all temperatures. Both the models are fit to the same experimental dataset.